# Investigation of Room Temperature Ferroelectricity and Ferrimagnetism in Multiferroic $Al_xFe_{2-x}O_3$ Epitaxial Thin Films

Badari Narayana Rao[1], Shintaro Yasui[1], Tsukasa Katayama[2], Ayako Taguchi[3,4], Hiroki Moriwake[3,4], Yosuke Hamasaki[5], Mitsuru Itoh[1]

1) Laboratory for Materials and Structures, Tokyo Institute of Technology, 4259 Nagatsuta, Midori, Yokohama 226-8503, Japan
2) Department of Chemistry, The University of Tokyo, Bunkyo-ku, Tokyo 112-0033, Japan
3) Nanostructures Research Laboratory, Japan Fine Ceramics Center, Atsuta-ku, Nagoya 456-8587, Japan
4) Center for Materials Research by Information Integration (CMI2), Research and Services Division of Materials Data and Integrated System (MaDIS), National Institute for Materials Science (NIMS), 1-2-1 Sengen, Tsukuba, Ibaraki 305-0047, Japan
5) Department of Applied Physics, National Defence Academy, Yokosuka 239-8686, Japan

**Abstract:** Multiferroic materials open up the possibility to design novel functionality in electronic devices, with low energy consumption. However, there are very few materials that show multiferroicity at room temperature, which is essential to be practically useful. $Al_xFe_{2-x}O_3$ (*x*-AFO) thin films, belonging to the $\kappa$-$Al_2O_3$ family are interesting because they show room temperature ferrimagnetism and have a polar crystal structure. However, it is difficult to realise its ferroelectric properties at room temperature, due to low resistivity of the films. In this work, we have deposited *x*-AFO ($0.5 \leq x \leq 1$) epitaxial thin films with low leakage, on $SrTiO_3$<111> substrates by Pulsed Laser Deposition. Magnetic measurements confirmed room temperature ferrimagnetism of the films, however the Curie temperature was found to be influenced by deposition conditions. First principle calculations suggested that ferroelectric domain switching occurs through shearing of in-plane oxygen layers, and predicted a high polarization value of 24 μC/cm$^2$. However, actual ferroelectric measurements showed the polarization to be two order less. Presence of multiple in-plane domains which oppose polarization switching of adjacent domains, was found to be the cause for the small observed polarization. Comparing dielectric relaxation studies and ferroelectric characterization showed that oxygen-vacancy defects assist domain wall motion, which in turn facilitates polarization switching.

## I. Introduction

Single-phase multiferroic materials have attracted considerable attention among scientists, due to strong drive in the industry towards device miniaturization and prospect of new functionalities with low energy consumption.[1–6] However, most of the known multiferroic materials have very low operational temperatures, thereby limiting their application.[7,8] Till date, only $BiFeO_3$ based multiferroic materials have shown promising properties with acceptable operational temperatures.[9,10] However, the antiferromagnetic nature of $BiFeO_3$ makes it less favourable for application. In addition, the high volatility of bismuth makes its fabrication difficult, which has led to inconsistency in the results obtained from different groups.[11] The $\kappa$-$Al_2O_3$-type family of oxides (e.g. $GaFeO_3$, $\varepsilon$-$Fe_2O_3$, $AlFeO_3$), are one of the alternative systems possessing both ferrimagnetism and ferroelectricity.[12–19] The $Ga_xFe_{2-x}O_3$ system was the first compound in this family to be discovered as multiferroic, as early as 1964.[20] However, unlike $GaFeO_3$, other compounds in this family are metastable phases which cannot be synthesized easily, and hence did not garner much interest. The recent advances in synthesis of nanoparticles and thin films have made stabilization of these phases possible. Multiferroic properties of these materials could be observed close to room temperature, and are currently being explored for various applications.[12,13,18–32] The ferrimagnetic nature of materials in this family make it advantageous over BiFeO$_3$, due to better magnetic properties.[33] Orthorhombic Al$_x$Fe$_{2-x}$O$_3$ ($x$-AFO) with space group *Pna*2$_1$, belongs to the same family of multiferroic oxides. This system is favourable, since it is made up of only 'Al' and 'Fe' cations, both of which are abundantly available, and are non-toxic in nature. Recently, thin films and nanoparticles of orthorhombic AlFeO$_3$ were successfully synthesized.[14,15,34,35] Hence, in the current work, we investigate the ferroelectric and magnetic properties of $x$-AFO epitaxial thin films deposited by pulsed laser deposition (PLD).

The orthorhombic structure of $x$-AFO is best described as that consisting of combination of hexagonal and cubic close-packing of oxygen ions. It contains one corner sharing tetrahedral site (Al1), one regular octahedral (Al2) and two heavily distorted octahedral sites (Fe1 and Fe2) which are edge shared [Fig. 1]. Though the cation sites are disordered in nature, the Al1 and Al2 sites are predominantly occupied by Al$^{3+}$ due to their smaller size, while Fe$^{3+}$ prefers the Fe1 and Fe2 sites.[14,33–35] The ferrimagnetism in $x$-AFO originates from the strong superexchange antiferromagnetic interactions between Fe ions,[14] where the Fe ion magnetic moment of Fe1 and Al1 sites are antiparallel to those at Fe2 and Al2 sites. In the case of $\varepsilon$-Fe$_2$O$_3$, it has been recently suggested that the spins of Fe$^{3+}$ in the Al1 site is non-collinear with respect to other sites, thereby inducing a larger magnetization in the system.[36] However, this effect may not be significant in the $x$-AFO system, since the occupancy of Fe$^{3+}$ in the Al1 site is small. A net magnetic moment in $x$-AFO mainly arises due to unequal distribution of Fe$^{3+}$ in the four cation sites.

While there are some research articles available on polycrystalline AlFeO$_3$ ceramics as well as thin films that have confirmed ferrimagnetism,[34,35,37] the evidence for piezoelectricity or ferroelectricity have not been very convincing.[13,38] Even the ferroelectricity in similar systems like GaFeO$_3$ and $\varepsilon$-Fe$_2$O$_3$ is puzzling, since the experimentally observed polarization values were considerably less than that predicted by ab-initio calculations.[17,36,39] Recently, Hamasaki *et al.* successfully synthesized epitaxial thin films of $x$-AFO on SrTiO$_3$(111) substrates.[15] However, high leakage currents in the films made direct ferroelectric measurements difficult, and only local information using piezoresponse force microscopy (PFM) could be obtained.[35] Due to this problem, very limited work on the ferroelectric and dielectric property measurements of $x$-AFO is available, thereby making such a study very interesting. The low resistance in thin films is generally due to high density of defects like oxygen-ion vacancies. The oxygen vacancies can be minimized either by tuning the deposition conditions or by suitable cation doping.[18,40–42] In the present work, we successfully optimized the deposition conditions to obtain $x$-AFO ($0.5 \leq x \leq 1$) films with low leakage current, thereby enabling ferroelectric and dielectric measurements. While first-principle calculations aided in understanding the mechanism of ferroelectric switching, the discrepancy in the polarization values obtained from theory and experiments is attributed to constraints posed by domains.

## II. Experiment
### A. Optimization of thin film deposition

$x$-AFO ($0.5 \leq x \leq 1$) films were deposited on STO(111) single crystal substrates by pulsed laser deposition (PLD) using fourth-harmonic wave of a Nd:YAG laser ($\lambda = 266$ nm) with repetition rate of 5 Hz. The films were also deposited on 0.5 wt% Nb-doped STO(111) (Nb:STO) conducting substrates for ferroelectric and capacitance measurements. As a PLD target, we used $x$-AFO ceramic pellets prepared by solid state synthesis (sintered at 1450°C for 14 hours). Several $x$-AFO films were deposited, and ideal conditions were identified by systematic variation of different parameters such as laser fluence (1 – 4 J/cm$^2$), oxygen partial pressure (10 mTorr – 500 mTorr), substrate temperature (650°C – 750°C) and annealing method. Single phase films were obtained in all the tested conditions, indicating a wide range of phase stability for the system. It was found that the deposition temperature is the critical parameter to obtain smooth films, while oxygen pressure ($P_{O_2}$) during deposition and annealing is important for obtaining single phase and controlling oxygen vacancies. A low deposition rate is favourable to obtain films with good electrical properties, which is controlled by laser fluence, $P_{O_2}$ and target-substrate distance. At higher laser fluence, it was noted that large number of droplets ejected from the target along with the plume, which then deposited on the film as amorphous macro particles, leading to poor quality films. A laser fluence of 1.6 J/cm$^2$ was found to be suitable for deposition of films of all compositions. The chamber pressure is an important parameter to control the shape of the plume, which improves the uniformity and smoothness of the film. An oxygen atmosphere helps to maintain the stoichiometry of the oxides by reducing oxygen vacancies in the film. $P_{O_2}$ of 100 – 300 mTorr was found to be ideal to obtain good quality films with good deposition rate. The substrate temperature controls the grain size and roughness of the films,[43] and a substrate temperature of about 710°C was found to be ideal for the film deposition. Annealing under high $P_{O_2}$ can decrease oxygen related defects in the film. However, optimum annealing temperature is important, since higher annealing temperature can lead to grain growth and increase in surface roughness. Annealing with $P_{O_2}$ of 100 Torr at 600°C for half an hour was found to be sufficient to obtain good films. After careful consideration, the following PLD conditions were used to deposit the films for further characterization: laser fluence of 1.6 J/cm$^2$, $P_{O_2}$ of 100 mTorr during deposition, substrate temperature of 710°C, and annealing with $P_{O_2}$ of 100 Torr at 600°C for half an hour. The film thickness was about 20 – 25 nm for all the compositions.

### B. Thin film characterization

The crystal structure of the films was analysed by high-resolution X-ray diffraction of Rigaku Smartlab using Cu-K$\alpha_1$ radiation. The ferroelectric measurements were carried out using the Precision Multiferroic II tester (Radiant Inc.). The dielectric measurements were carried out using an LCR meter (Agilent, 4284A), while the samples were loaded inside a Physical Property Measurement System (PPMS, Quantum Design Inc.). Out of plane piezoresponse force microscopy measurements were carried out using a frequency tracking DART mode of MFP-3D Asylum Research microscope. For all electric measurements, Pt top electrode (100 μm diameter) was deposited on the films by electron beam evaporation. While top-top electrode configuration was used for ferroelectric measurements,[44] the rest of the electrical measurements used Pt as top electrode, and the Nb:STO substrate as the bottom electrode. The in-plane magnetizations of the films were measured using a superconducting quantum interference device (SQUID) magnetometer (Quantum Design Co. MPMS XL).

### C. First principles calculation method

The *Ab initio* calculations were performed by the projector-augmented wave (PAW) method within the GGA+U formalism[45] and the framework of density functional theory (DFT)[46,47], as implemented in the VASP code[48,49]. The exchange-correlation interactions were treated by the generalized gradient approximation (GGA-PBE)[50]. The on-site Coulomb repulsion was treated at the GGA+$U$ level[51]. We adopted the Hubbard effective $U_{\text{eff}}$ = 4.0 eV only for the Fe-3$d$ electrons. For the PAW potentials, the electronic configurations 3$d^{10}$4$s^2$ for Fe, 3$s^2$3$p^1$ for Al and 2$s^2$2$p^6$ for O were explicitly treated as valence electrons. The plane wave expansion up to 600 eV was adapted. A $k$-point mesh of 4×2×2 within the 40-atom unit cell was used for Brillouin zone sampling of primitive cells, which was based on the Monkhorst-Pack scheme[52]. The lattice constants and internal atomic coordinates were considered fully optimized once the residual Hellmann-Feynman (HF) forces were less than $1.0\times10^{-2}$ eV/Å. The activation energy for this switching was determined using the nudged elastic band (NEB) method[53]. The polarization values were determined by Berry's phase[54] method implemented in the ABINIT code[55].

## III. Results
### A. Structural characterization

Fig. 2(a) shows the out-of-plane $2\theta\text{-}\theta$ XRD scan of 0.5-AFO film grown on STO(111) substrates, which shows that a single phase is successfully obtained. Since only 00$l$ peaks of the film are observed, it is clear that the film growth is $c$-axis-oriented. The thickness fringes observed in the 004 peak (inset of Fig. 2(a)) indicate smooth film, and was observed for all the compositions studied. The in-plane crystal-domain orientations of the films were evaluated using φ-scan about the $x$-AFO{201} and STO{110} diffraction peaks [Fig. 2(b)]. The film showed six-fold in-plane symmetry of the {201} peak, indicating three types of in-plane domains, where each [100]$_{Film}$ direction is parallel to the [11-2]$_{STO}$, [1-21]$_{STO}$, or [-211]$_{STO}$ direction, as illustrated in Fig. 2(c). These results are consistent with previous reports of AFO and GaFeO$_3$ based films on STO(111) substrates.[15,56] Fig. 2(d) shows the variation of the lattice parameters obtained from in-plane and out-of-plane X-ray diffraction, as a function of composition. A decrease in unit cell volume is observed with increasing Al-content (Fig. 2(e)), which is consistent with Vegard's law, as the ionic radii of $Al^{3+}$ is smaller than $Fe^{3+}$.

### B. Magnetic properties

Figure 3 shows the magnetic properties of $x$-AFO films. Fig 3(a) shows the field cooled temperature dependence of magnetization (*MT*) for different compositions, indicating that the Curie temperature for all the films is above 400 K. Since the maximum operational temperature of the SQUID was 400K, the exact Curie temperature could not be determined. However, the magnetic measurements on films grown at 300 mTorr oxygen pressure showed relatively lower Curie temperatures [Fig. S1], indicating that Curie point can be tuned by varying oxygen pressure. From Fig. [S1], it is also clear that the magnetic Curie temperature decreases with increasing $x$, similar to other reports.[19,35] Figure 4(b) shows the room temperature magnetization vs. magnetic field hysteresis (*MH*) plots for $x$ = 0.5, 0.8 and 0.9, revealing their ferrimagnetic nature. The inset in fig. 4(b) shows the zoomed plot of the *MH* curve, indicating higher coercive field for lower value of $x$. Figure 4(c) shows the actual variation of the magnetic coercive field and the saturation magnetization as a function of composition. While the coercive field continuously decreased with increasing $x$, the saturation magnetization was maximum at 0.8-AFO. Pure ε-Fe$_2$O$_3$ ($x$ = 0), has a very high coercive field due to the strong hybridization of Fe $3d^5$ at the Fe2 site with O $2p$ orbital, resulting from large spin-orbit interaction[57]. The decrease in coercive field with increasing $x$ (decreasing Fe concentration) must be due to weakening of this phenomenon, since the system is moving away from ε-Fe$_2$O$_3$. The reason for maxima in saturation magnetization at $x$ = 0.8 can be attributed to the differential occupation of $Al^{3+}$ in each of the four cation sites. This can be explained using figure 3(d), which shows an illustration of the magnetic moments of the four cation sites. When $x$ = 0 (pure ε-Fe$_2$O$_3$), all the sites are completely occupied by $Fe^{3+}$ ions. In this situation, sites Fe1 and Al1 have their spins aligned in one direction and sites Fe2 and Al2 have the spins aligned in the opposite direction. The non-collinear magnetic moment in the tetrahedral site results in a non-zero magnetic moment.[36] As $x$ increases, for lower values of $x$ ($x$ < 0.8), the $Al^{3+}$ preferentially occupy the Al1 sites. Since Al1 and Fe1 sites are antiparallel to Al2 and Fe2 sites, the net magnetic moment is given by the algebraic sum of moments from all the sites: $M_{Al2} + M_{Fe2} - M_{Al1} - M_{Fe1}$. Hence, with increasing $x$, the magnetic moment contribution from Al1 site decreases, thereby increasing the net magnetic moment. However, as $x$ increases further ($x > 0.8$), Al ions begin to occupy the Al2 sites also, consequently decreasing the net magnetic moment.

### C. Ferroelectric Properties

While the $x$-AFO system has a non-centrosymmetric structure, its ferroelectricity has never been adequately verified. The films are prone to leakage, which makes ferroelectric measurements difficult. Though we could observe domain switching as well as butterfly amplitude loop in PFM measurements (Fig. S2), it is not sufficient to prove its ferroelectric nature. This is because non-ferroelectric surfaces are also known to show contrast in PFM measurements under certain conditions.[58–61] Hence, we focussed on direct ferroelectric measurements, which was possible by obtaining films with improved leakage properties. Fig. 4 shows the polarization vs. electric-field (*PE*) hysteresis loops for different compositions of the films. It can be seen that all compositions showed good hysteresis loops, and domain switching is confirmed by peak in the current vs. electric-field (*IE*) plot, corresponding to the coercive field. However, it was observed that the hysteresis loops from these films do not saturate until the breakdown field, as shown in Fig. 5a-b. Figure 5a shows the plot of *PE* hysteresis measured with increasing maximum electric field for $x = 1$, and fig. 5b confirms that both the remnant polarization as well as the coercive field do not saturate with the electric field. This behaviour is different compared to conventional ferroelectrics, which show sudden anomaly in polarization above coercive field, and saturates at higher fields. Absence of sudden jump in remnant polarization of our films indicates that the polarization switching mechanism may be different as compared to conventional ferroelectrics. A continuous increase in the polarization with increasing electric field indicates major contribution from paraelectric effect, which ideally has a linear relationship with electric field. The presence of the paraelectric component in the polarization of our film was further confirmed by the fact that the shape of the *PE* loops became slimmer with decreasing frequency [Fig. S3]. However, the peak in the *IE* curves indicate polarization switching, which means that spontaneous polarization due to ferroelectricity is also present.

To further investigate ferroelectricity in $x$-AFO films, the paraelectric and ferroelectric components of polarization were separated out by the Positive Up Negative Down (PUND) measurement technique[62]. Fig. 5c and 5d shows comparison of the remnant polarization obtained from *PE* hysteresis measurements and PUND measurement for 0.5-AFO film. We can see that the actual remnant polarization as obtained from PUND measurement is smaller than that determined by *PE* loops. Similar results were obtained for other compositions of $x$-AFO as well. This proves our earlier proposition that a large component of the polarization arises from the paraelectric component. It must be noted that, the *PE* loop and PUND measurement results were susceptible to minor unavoidable differences in deposition conditions. Even two films of same composition, which were deposited separately, showed slight variations in their polarization values. As a result, we did not notice any significant composition dependence of the ferroelectric properties.

### D. Dielectric Properties

Figure 6 shows the dielectric data obtained for 0.5-AFO film between 2 to 350 K in the frequency range from 100 Hz to 1 MHz. Frequency dispersion is observed over a wide temperature range, beginning at about 150 K and continuing well above room temperature. Such large dispersion is often attributed to motion of ferroelectric domain boundaries.[63] All compositions studied ($0.5 \leq x \leq 1$) showed similar behavior, with small shifts in the temperatures corresponding to the peaks in dielectric loss

(fig. 7 (a-c)). Figure 7d and 7e shows the temperature dependence of imaginary part of dielectric constant as a function of frequency, for $x = 0.5$ and 1 respectively. It can be seen from the figure that the dispersion occurs over a wide range of frequency. The origin of the relaxation in the films were further analyzed by modeling the frequency dependence of the peak positions in the dielectric loss curves, using an Arrehenius relation (Fig. 8(f)): $F = F_o exp\left[\frac{-E_a}{k_B T}\right]$, where $F$ is the measuring frequency, $E_a$ is the activation energy, and $F_o$ is the attempt jump frequency. An activation energy of about 0.37 eV, and $F_o$ of the order of $10^{10}$ Hz were obtained for all the compositions (Table 1). Since the mobility of Al and Fe ions are negligible at such low temperatures, we associate the relaxation process to be dominated by oxygen vacancies. These oxygen vacancies generally aggregate near domain boundaries, and since TEM observations showed the domain size to be very small (5-10 nm)[35], the defect density in the film could be very high. Thus oxygen vacancies can contribute significantly to the electrical properties of the film. We propose the ferroelectric domain motion to be assisted by electron hopping through the $Fe^{2+}$-$V^{\bullet}_o$-$Fe^{3+}$ route, where $V^{\bullet}_o$ denotes electron-trapped oxygen vacancy following the Kröger–Vink notation. This is very likely since oxygen has a rather small first ionization energy (0.1 eV)[64]. Similar order of activation energy for electron hopping through the oxygen vacancy has also been observed by Ke et al. for (La,Mg) substituted $BiFeO_3$[65], by Ikeda et al. for $LuFe_2O_4$[63], and by Katayama et al. for $Ga_xFe_{2-x}O_3$[19].

Since the dielectric relaxation is found to be associated with domain motion, which in turn is assisted by $Fe^{2+}$-$V^{\bullet}_o$-$Fe^{3+}$ hopping, we tried to correlate the results from dielectric analysis and ferroelectric measurements. It can be seen from fig. 8 that at any particular temperature, a clear *PE* hysteresis is obtained only at frequencies close to the relaxation frequency observed in the dielectric data. At lower temperatuers, the relaxation is observed at lower frequencies, and consequently, a good *PE* hysteresis loop is also obtained at the same frequency. From the above observation, it is clear that polarization switching is intricately correlated to oxygen vacancy defects, which are usually found in the vicinity of domain boundaries. Upon application of electric field, the local electric dipole formed by these defects trigger the actual ferroelectric domain switching process (shown in Fig. 9). Since the defects are most mobile at their relaxation frequency, even the polarization response is best observed at this frequency (Fig. 8).

### E. First-Principles Calculation

Theoretically determined activation energies and polarization switching mechanisms of *x*-AFO are discussed based on *ab initio* calculations performed on *κ*-$Al_2O_3$ and *ε*-$Fe_2O_3$. One possible mechanism of polarization switching of *κ*-$Al_2O_3$ type *x*-AFO is via an intermediate non-polar centrosymmetric state[17,66]. Earlier, Stoeffler et al. calculated the activation energy and net polarization of isostructural $GaFeO_3$ by considering a *Pnna* space group as the intermediate state[17]. However, the reported activation energy for the polarization switching was 0.5 eV, which is much larger than that seen in conventional ferroelectric compounds (e.g. $BaTiO_3$ – 0.02 eV[67], $PbTiO_3$ – 0.03 eV[68]). Xu et al. suggested an alternative centrosymmetric space group *Pbcn*, which gave a much lower activation energy for polarization in *ε*-$Fe_2O_3$[36]. Hence, we considered the *Pbcn* space group as the non-polar polarized structure for our calculation. Figure 9(a - e) show the schematic of transition from a negatively polarized structure to centrosymmetric, and then to a positively polarized structure. The calculation yielded activation energies for polarization switching of 0.088 and 0.155 eV/f.u for *ε*-$Fe_2O_3$ and *κ*-$Al_2O_3$ respectively (Fig. 10). We can expect the activation energies for the intermediate *x*-AFO structures also to be of similar order. These values are fairly small and acceptable, compared to the high value previously reported for $GaFeO_3$[17,66]. During the polarization reversal process, the polarization switches from $-P_s$ to $+P_s$, while smoothly passing through zero (Fig. 9(e & h)). Viewing the structure along the *b*-axis clearly explains the polarization switching mechanism (Fig 9(f-j)). Close-packed oxygen layers in corundum layers keep their octahedral shape during the switching. However, oxygens above and below the corundum layers shift along *a*-axis, in opposite directions relative to each other. This shearing motion of oxygen layers induces a coordination switching of cations Fe1 and Al1 sites. An originally tetrahedral(octahedral) Al1(Al2) site turns into octahedral(tetrahedral) Al2(Al1) site after the polarization switching. This mechanism is quite different from conventional ferroelectric perovskite oxides, where cations and anions move in opposite directions in a linear manner (Slater mode[69]).

By using the Berry's phase approach[54], the polarization of $Al_2O_3$ and $\varepsilon$-$Fe_2O_3$ was calculated to be about 26 $\mu C/cm^2$ and 21 $\mu C/cm^2$ respectively. Since there is no structure change in the substituted *x*-AFO series, their theoretical polarization values will also lie in between 21 - 26 $\mu C/cm^2$. While this value of polarization is comparable to theoretical values reported for other isostructural compounds like $GaFeO_3$ and $\varepsilon$-$Fe_2O_3$,[17,36,39,66] it is about two orders of magnitude larger than that observed experimentally. Similar ambiguity is observed in $GaFeO_3$ based films as well, and the exact reason for this is not yet known. We speculate that the multi-domain structure of the thin films obstruct complete polarization reversal, and hence the actual polarization is considerably lesser than that predicted.

### IV. Discussion

Among all the known multferroics in the world, the $BiFeO_3$ system has attracted the largest attention, due to its high Néel's temperature and large ferroelectric polarization. However, problems like the antiferromagnetic ordering of $BiFeO_3$ and volatility of Bi during fabrication makes it unattractive for magnetic or magnetoelectric applications. Hence, it is necessary to identify other potential multiferroic materials, thereby giving more flexibility for the electronic industry. $\kappa$-$Al_2O_3$ type ferrites like $\varepsilon$-$Fe_2O_3$, $GaFeO_3$ and $AlFeO_3$ have recently been identified to be promising multiferroics.[18,19,23,26] Especially, the ferrimagnetic nature of these ferrites, which can be stabilized above room temperature, is a prime advantage over $BiFeO_3$. While the large coercive field and magnetic anisotropy of $\varepsilon$-$Fe_2O_3$ has already made it interesting for high-frequency millimeter wave absorption,[23] the research on ferroelectric and multiferroic properties of these ferrites is still in its nascent stage. Recently, Katayama *et. al.* showed that the properties of $\kappa$-$Al_2O_3$ type $GaFeO_3$ can be tuned by suitable cation substitution, to obtain excellent ferroelectric and multiferroic properties at room temperature.[18] The *x*-AFO system is made of only Al and Fe ions, both of which are abundantly available and are non-toxic in nature. Hence, it can be a potential gamechanger in the electronic industry, if good ferroelectric and magnetic properties can be established in this system. The magnetic properties are easier to be tuned, and a Curie temperature above room temperature could be established. The Curie temperature can be tuned either by cation substitution or by varying oxygen vacancy concentration. Jaffe *et al.* suggested that *n*-type carriers present due to oxygen vacancy in semiconducting ferromagnets help mediate magnetic interactions between spins[70]. A similar phenomena may be effecting the *x*-AFO system as well, as a consequence of which the Curie temperature is influenced by oxygen vacancy concentration[71].

Though theoretical predictions state much larger polarization (~20-24 $\mu C/cm^2$), the actual value is about two order less. This ambiguity has been observed for other systems in the family ($GaFeO_3$, $\varepsilon$-$Fe_2O_3$) as well.[17,36] We suggest that the existence of multiple in-plane domains in the film could be the reason behind reduced polarization. As we have shown using first-principle calculation that the domain switching along *c*-axis (out-of-plane) takes place by shearing of oxygen layers along *a*-axis, presence of multiple in-plane domains will make such a shearing very difficult. The X-ray diffraction

$\phi$-scans of the films in Fig. 2b clearly show presence of three types of crystal domains, adhering to the 3-fold symmetry of the STO(111) substrate surface. Any attempt of shearing of oxygen from domain **1** will be constricted by domains **2** and **3**, and likewise, shearing in domains **2** and **3** will also be restricted (Fig. 11). Hence, when electric field is applied, the domains can orient only to a small extent, and they tend to go back to the original position upon removal of the electric field. This model can simultaneously explain the low polarization values, the large paraelectric contribution to the polarization, as well as the correlation of polarization to dielectric relaxation of the films. If we can grow films with single domain, then it is likely that polarization values close to that of theoretical calculations can be obtained. An extensive work on domain engineering is required to obtain single domain films of $x$-AFO. Preliminary work by Katayama *et al.* on $Ga_{0.6}Fe_{1.4}O_3$ films showed that STO(111) yields the minimum number of in-plane domains compared to several other substrates.[56] Also, using STO(100) and STO(110) oriented substrates yielded six in-plane domains, which is double of that obtained in STO(111). Hence, among all the substrates studied so far, STO(111) yields the least number of in-plane domain types. Further research on more substrates and deposition conditions is required to achieve single domain films.

Xu *et al.* have shown by theoretical calculations that cation size is an important factor in stabilizing the ferroelectric phase.[36] A decrease in the cation size stabilizes the ferroelectric phase, and since the radius of $Al^{3+}$ (0.535 Å) is smaller than that of $Fe^{3+}$ (0.55Å and 0.645Å in high and low spin state respectively), increasing $x$ should improve the ferroelectric property of the system. However, in the present study, we could not establish any composition dependence of ferroelectric properties. Since the polarization reversal in these systems occurs in an indirect manner, many intricate parameters like oxygen vacancy concentration, occupancy of each cation site, defect population, etc, may supersede the effect due to cation size. Nevertheless, room temperature ferroelectricity has been clearly demonstrated in the $x$-AFO films. The system also shows dielectric dispersion, with the activation energy about 0.37 eV, corresponding to hopping between localized charge carriers. In comparison to $BiFeO_3$, these systems have better magnetic properties, and hence promise to be more beneficial for multiferroic applications.

## V.  Conclusion

Room temperature ferroelectricity and ferrimagnetism has been established in the $x$-AFO system. The ferroelectric response of the films is highly frequency dependent, and the deposition conditions have remarkable influence on the nature of the *PE* hysteresis loops. PUND measurements proved to be better at providing reliable remnant polarization values. The low polarization in the films could be attributed to constraints posed by multiple in-plane domains. However, we hope that the current work will lead to further development in fabrication of single domain films, which can then yield polarization values close to the theoretical ones. The magnetic measurements were consistent with other works, and the Curie temperature and coercive field were found to decrease with increasing $x$. The room temperature magnetism and ferroelectricity of the $x$-AFO system, which comprises of inexpensive and nontoxic raw materials, makes this system promising for multiferroic applications.

**Acknowledgements**

B.N.R acknowledges fellowship support by JSPS(P17079). This work was partly supported by JSPS KAKENHI Grants-in-Aid for challenging Research (Pioneering) (M.I., 1706420), (Exploratory) (Sh.Y., 18K19126), and MEXT Elements Strategy Initiative to form Core Research Centre, Collaborative Research Project of Laboratory for Materials and Structures, Institute of Innovative Research, Tokyo Institute of Technology. H.M acknowledges "Materials research by Information

Integration" Initiative (MI2I) project of the Support Program for Starting Up Innovation Hub from Japan Science and Technology Agency (JST).

**Figures**

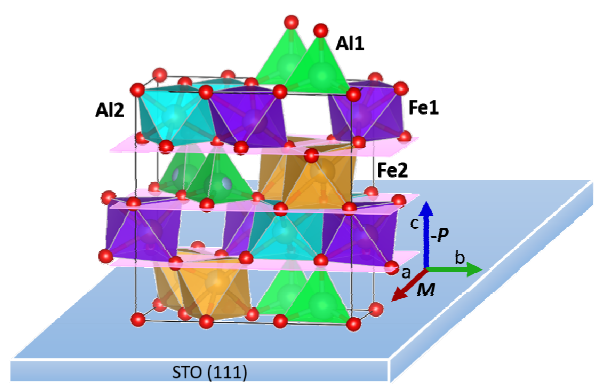

**Figure 1**: Crystal structure model of orthorhombic $x$-AFO with space group $Pna2_1$. Al1 indicates the tetrahedral cation site, whereas Al2, Fe1 and Fe2 indicate the octahedral cation site. $P$ and $M$ indicate the direction of ferroelectric polarization and magnetization respectively.

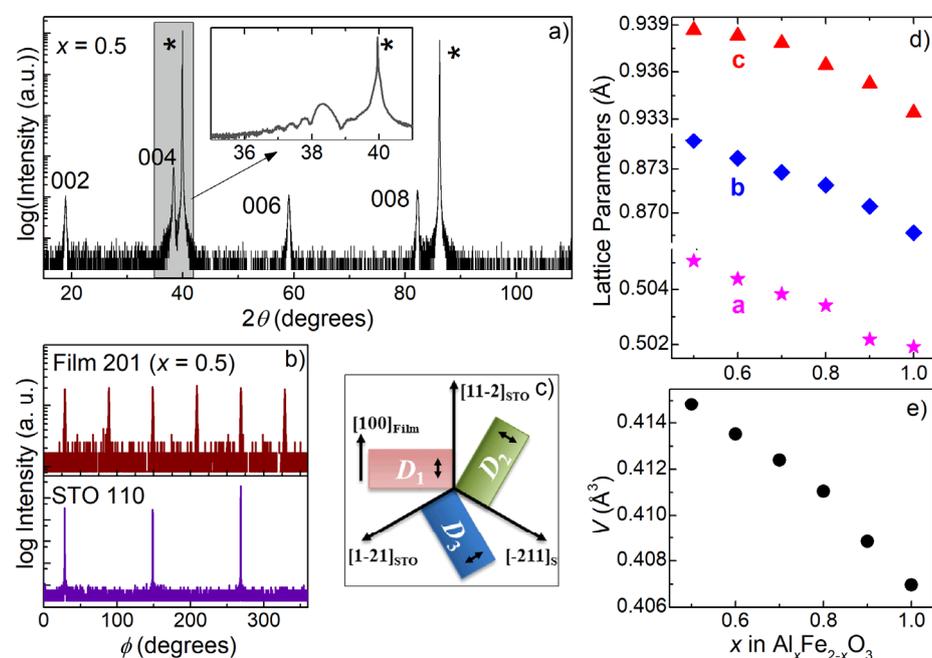

**Figure 2**: a) out-of-plane XRD pattern of 0.5-AFO film (* indicates substrate peaks). Inset shows the expanded view of 004 peak. b) phi scans of AFO film about 201 reflection and STO(111) substrate about 110 reflection. c) schematic of orientation relationship between film domains and the substrate. (d) Variation of lattice parameters as a function of $x$. e) Unit cell volume as a function of $x$.

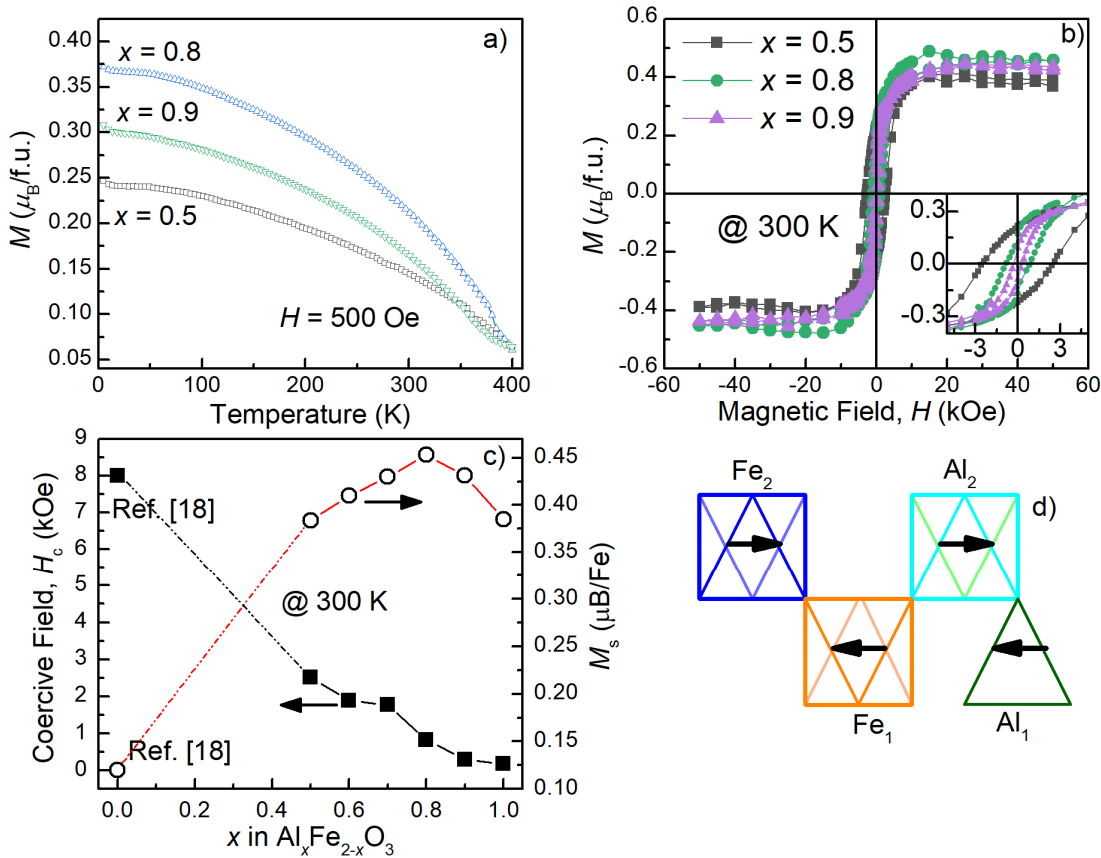

**Figure 3.** a) Field cooled magnetization vs temperature plot for various compositions (*MT*), measured with a constant magnetic field of 500 Oe. b) Room temperature magnetization vs magnetic field (*MH*) for different compositions, inset shows the zoomed version of the same plot to highlight the differences in the coercive field. c) Compositional variation of saturation magnetization ($M_s$) and coercive field ($H_c$), at room temperature. d) Schematic of the direction of magnetic moments of $Fe^{3+}$ ions at each site.

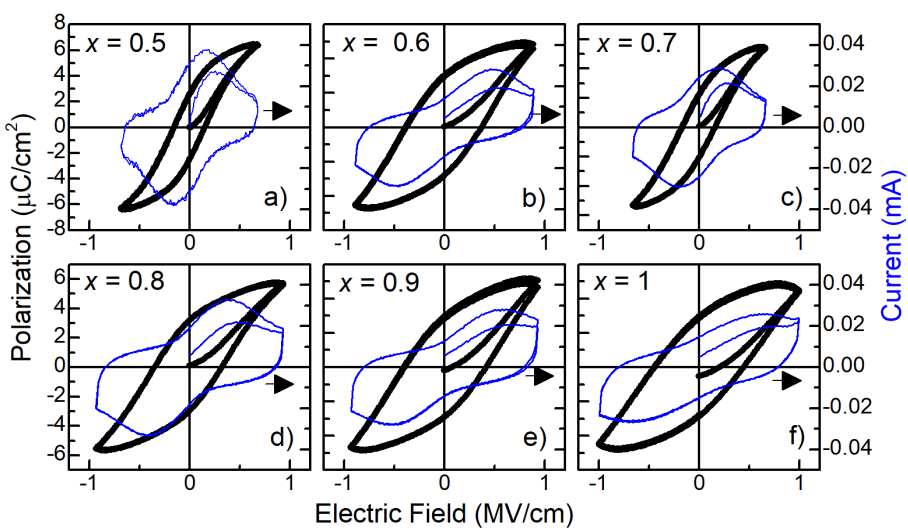

**Figure 4.** (a-f) Polarization vs Electric field (*PE*) and current dependence of electric field (*IE*) for various compositions of the film, collected at 10 kHz.

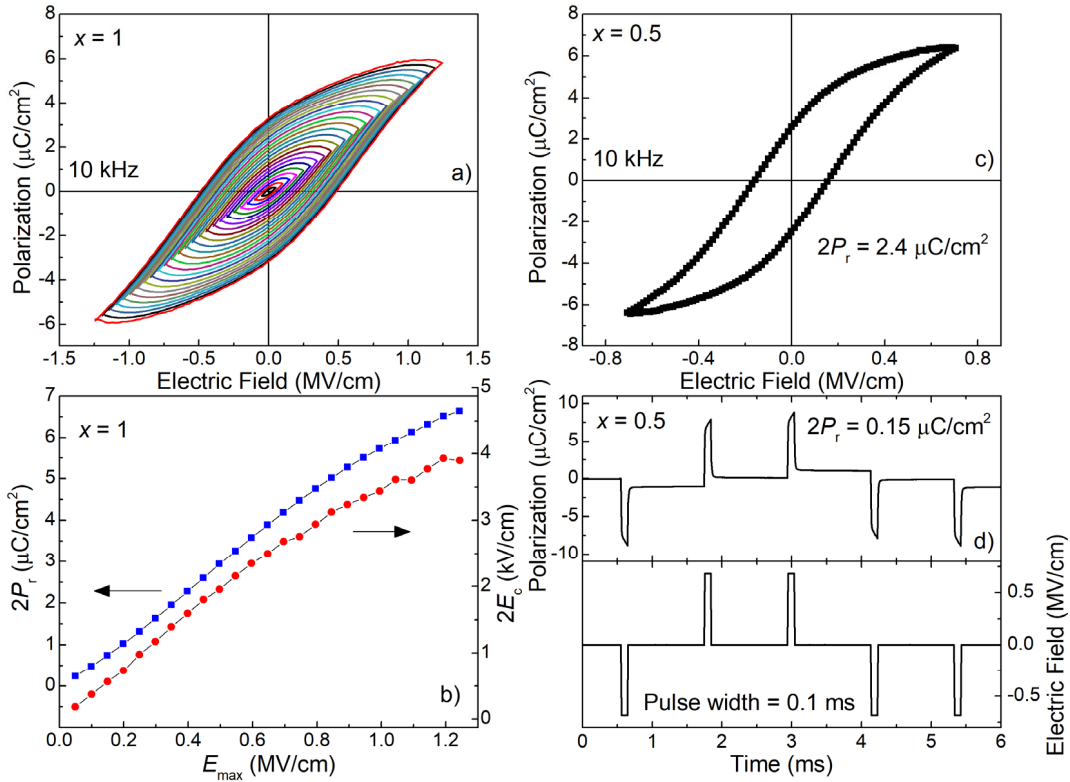

**Figure 5.** (a) PE hysteresis loops of 1-AFO at 10 kHz, with increasing maximum electric fields. (b) Variation of remnant polarization and coercive field of 1-AFO as a function of maximum electric field, as obtained from (a). (c) Remnant polarization of 0.5-AFO as calculated from *PE* hysteresis. (d) Remnant polarization of 0.5-AFO as calculated from PUND measurement.

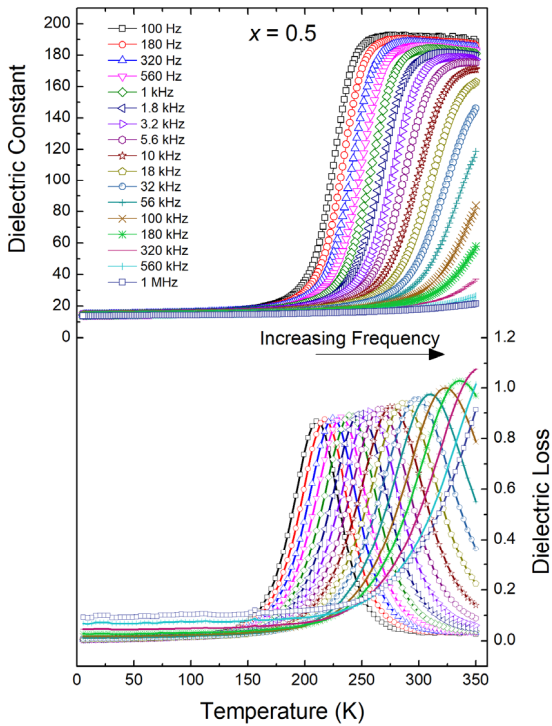

**Figure 6.** a) Dielectric constant and dielectric loss as a function of temperature and frequency, for 0.5-AFO

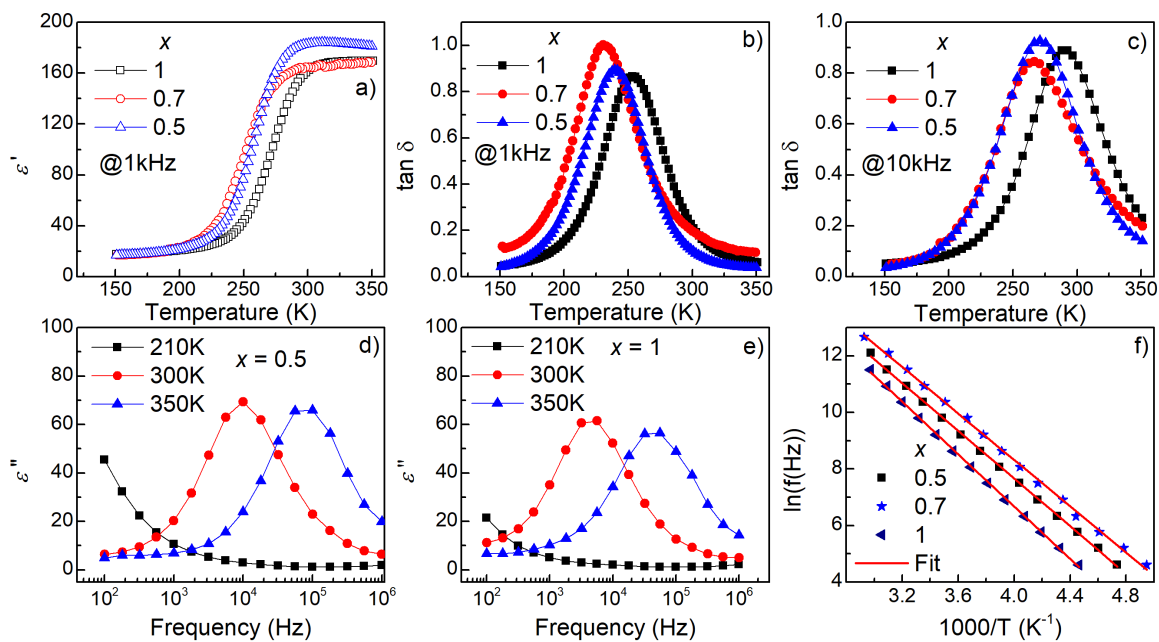

**Figure 7.** (a) Variation of real part of dielectric constant with temperature. (b,c) Variation of loss tangent with temperature at 1 kHz and 10 kHz respectively. (d,e) variation of imaginary part of dielectric constant with frequency for $x = 0.5$ and 1 respectively, clearly depicting the frequency dispersion. (f) Plot of ln(Peak Frequency) vs inverse temperature, to show that the system follows Arrhenius relation.

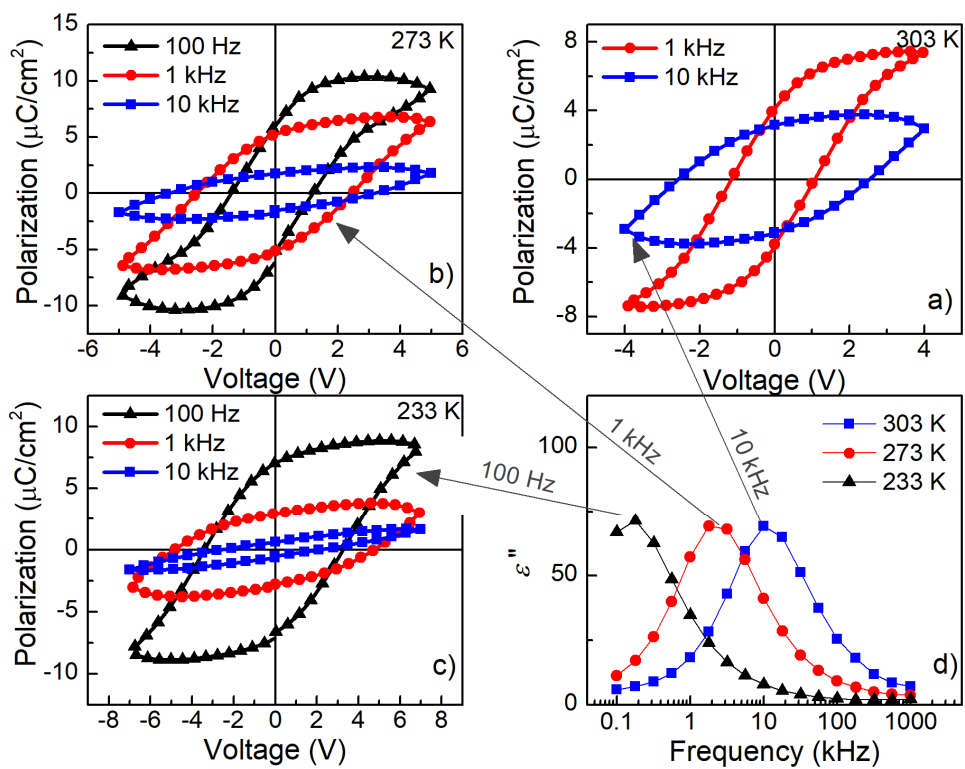

**Figure 8.** Comparison of dielectric plot and *PE*-hysteresis at different temperature and frequencies for $x = 0.5$.

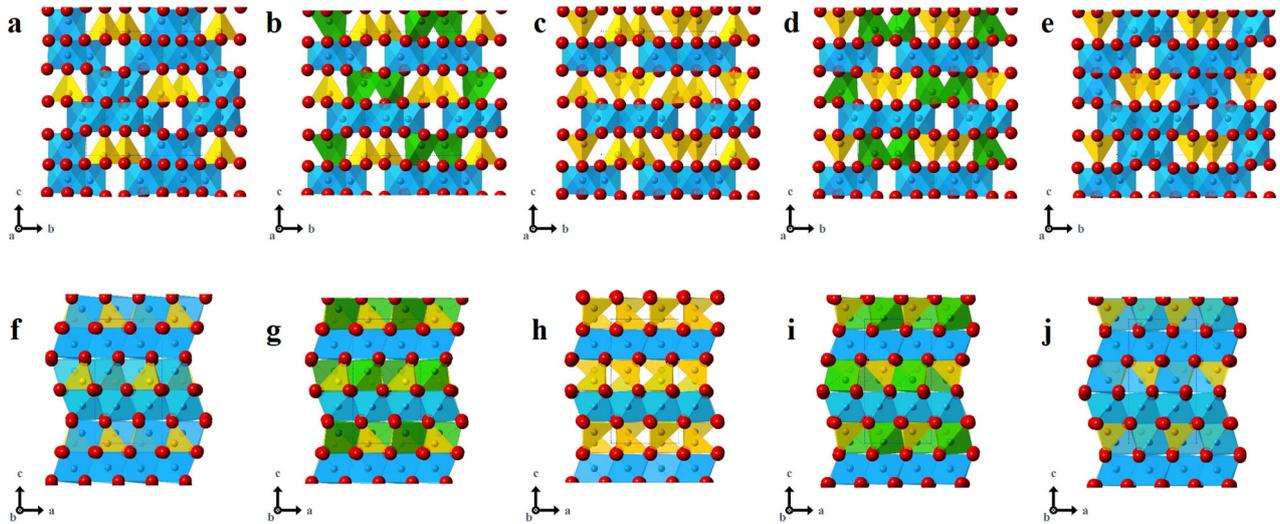

**Figure 9.** Illustration of structural changes upon polarization reversal. (a-e) structure viewed along *a*-axis, (f-j) structure viewed along *b*-axis. Yellow, green, and blue shapes indicate tetrahedral, pentahedra, and octahedra respectively.

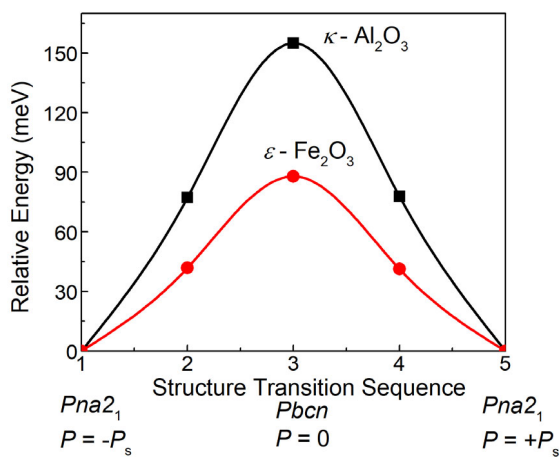

**Figure 10.** Variation of relative energies of $\kappa$-$Al_2O_3$ and $\varepsilon$-$Fe_2O_3$ during polarization switching through an intermediate centrosymmetric structure $Pbcn$.

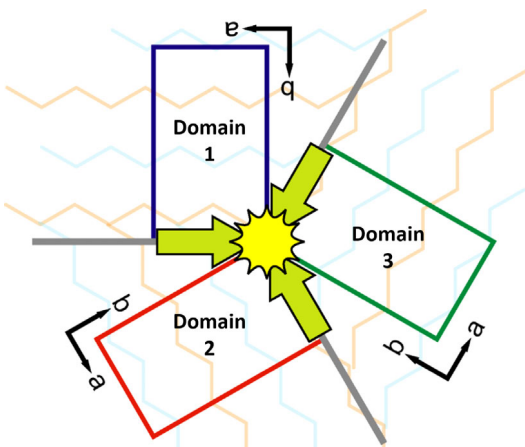

**Figure 11.** Illustration of constriction of the shearing of the oxygen layers, in the multi-domain structure.

**Table 1.** Activation energy ($E_a$) and attempt jump frequency ($F_o$) for various compositions of $x$-AFO, obtained by Arrhenius fitting of temperature dependent dielectric loss data.

| Composition | $E_a$ (eV) | $F_o$ (Hz) x $10^{10}$ |
|---|---|---|
| 0.5-AFO | 0.363 | 4.51 |
| 0.6-AFO | 0.378 | 7.76 |
| 0.7-AFO | 0.353 | 5.52 |
| 0.8-AFO | 0.402 | 11.03 |
| 1-AFO | 0.399 | 8.75 |

# Supporting Information

"Investigation of Room Temperature Ferroelectricity and Ferrimagnetism in Multiferroic Al$_x$Fe$_{2-x}$O$_3$ Epitaxial Thin Films"

Badari Narayana Rao[1], Shintaro Yasui[1], Tsukasa Katayama[2], Ayako Taguchi[3,4], Hiroki Moriwake[3,4], Yosuke Hamasaki[5], Mitsuru Itoh[1]

1) Laboratory for Materials and Structures, Tokyo Institute of Technology, 4259 Nagatsuta, Midori, Yokohama 226-8503, Japan
2) Department of Chemistry, The University of Tokyo, Bunkyo-ku, Tokyo 112-0033, Japan
3) Nanostructures Research Laboratory, Japan Fine Ceramics Center, Atsuta-ku, Nagoya 456-8587, Japan
4) Center for Materials Research by Information Integration (CMI2), Research and Services Division of Materials Data and Integrated System (MaDIS), National Institute for Materials Science (NIMS), 1-2-1 Sengen, Tsukuba, Ibaraki 305-0047, Japan
5) Department of Applied Physics, National Defence Academy, Yokosuka 239-8686, Japan

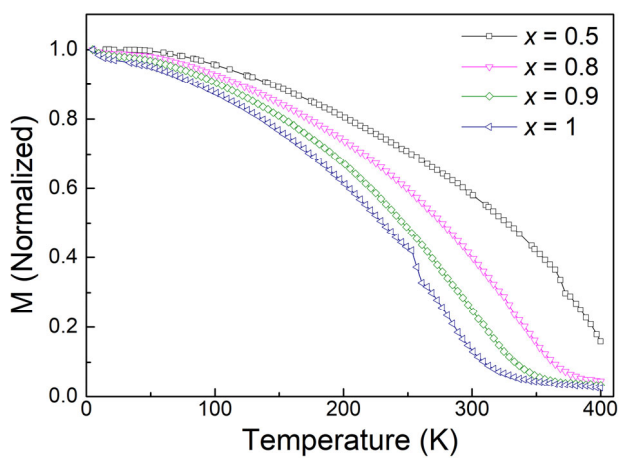

**Figure S1.** Normalized Magnetization vs. Temperature plot for films grown at 300 mTorr PO$_2$, showing lower Curie temperature than in Fig. 3, as well as decrease in Curie temperature with increasing $x$.

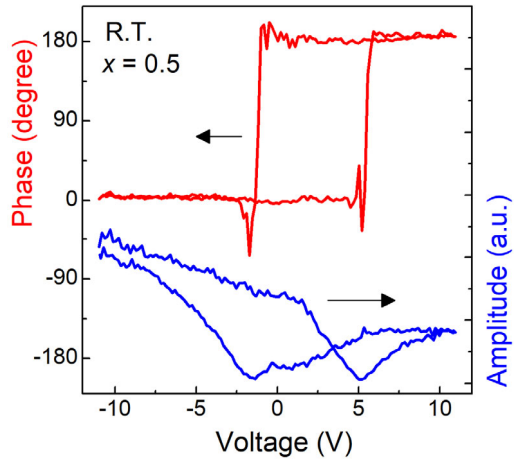

**Figure S2.** The phase and amplitude curves of 0.5-AFO films, obtained from piezoresponse force microscopy, clearly depicting domain switching.

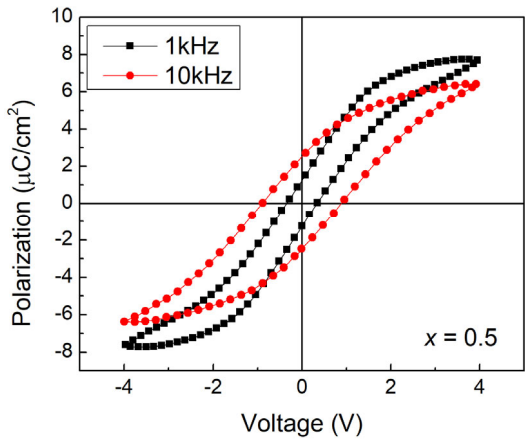

**Figure S3:** Frequency dependence of PE loop, for $x = 0.5$, showing slimming of the PE loop with decrease in frequency.